\documentstyle[aps,preprint]{revtex}
\tightenlines
\begin{document}

\begin{titlepage}
\title{On the binding energy of double  $\Lambda$ hypernuclei in the 
relativistic mean field theory.}

\vspace{6pt}

\author{
S. Marcos\\
Departamento de F\'{\i}sica Moderna\\
Universidad de Cantabria, 39005 Santander, Spain, \\[3mm]
R.J. Lombard\\ 
Division de Physique Th\'eorique \thanks{Unit\'e de Recherche des
Universit\'es Paris 11 et Paris 6 Associ\'ee au CNRS }, \\
Institut de Physique Nucl\'eaire\\
91406 Orsay Cedex, France,\\ [3mm]
and \\ [3mm]
J. Mare\v{s}\\
Nuclear Physics Institute, 250 68
\v{R}e\v{z}, Czech Republic\\
}

\vspace{6pt}
\maketitle

\begin{abstract}
We calculate the binding energy of two $\Lambda$ hyperons bound to a nuclear 
core within the relativistic mean field theory.
The starting point is a two-body relativistic equation of the Breit type
suggested by the RMFT, and corrected for the two-particle
interaction. 
We evaluate the 2 $\Lambda$ correlation energy and estimate the contribution
of the  $\sigma^*$ and $\Phi$ mesons, acting solely between hyperons,  
to the bond energy  $\Delta B_{\Lambda\Lambda}$ of $^6_{\Lambda\Lambda}He$, 
 $^{10}_{\Lambda\Lambda}Be$ and $^{13}_{\Lambda\Lambda}B$. 
Predictions of the  $\Delta B_{\Lambda\Lambda}$ A dependence are made for
heavier  $\Lambda$-hypernuclei.
\end{abstract}\vspace{1cm}
\end{titlepage}

\section{Introduction.}

Double $\Lambda$ hypernuclei, nuclear systems containing two $\Lambda$
hyperons, have been retaining much interest \cite{fran}.
As the only observed example of a multiply strange system they
give us unique opportunity to study $\Lambda\Lambda$ interaction in
nuclear medium and to
test existing models of the baryon-baryon interaction.
Moreover, studies of $\Lambda\Lambda$ hypernuclei are closely related
to searches for $S=-2$ dibaryon, known as the H particle \cite{jaff}.
Up to now, only a few events have been identified   
($^6_{\Lambda\Lambda}He$, $^{10}_{\Lambda\Lambda}Be$,
$^{13}_{\Lambda\Lambda}B$) \cite{dal,aoki,dov} indicating a strong 
attractive $\Lambda\Lambda$ interaction. The analysis of the data yields 
the $\Lambda\Lambda$ bond energy $\Delta B_{\Lambda\Lambda}
\approx 4-5$~MeV, where
\begin{equation}
\Delta B_{\Lambda\Lambda}=
B_{\Lambda\Lambda}(^A_{\Lambda\Lambda}Z) - 2 B_{\Lambda}(^{A-1}_{\Lambda}Z)
= 2 M(^{A-1}_{\Lambda}Z)
- M(^{A}_{\Lambda\Lambda}Z) - M(^{A-2}Z)\;\;\; ,
\end{equation}
 $B_{\Lambda}$ and $B_{\Lambda\Lambda}$ being the binding energy of
$\Lambda$ and a pair of $\Lambda$'s, respectively.

Calculations of $B_{\Lambda\Lambda}$ have been performed in various
non-relativistic approaches \cite{ban,ike,yam,him,bod,lan}, by using effective interactions or
G-matrices together with cluster or three-body models (a fair list of 
early works is given in [8]. In particular,
the D model of the Nijmegen $\Lambda\Lambda$ interactions was shown to
yield results in good agreement with experiments.

The purpose of the present work is to investigate $\Delta B_{\Lambda\Lambda}$
within the relativistic mean field (RMF) theory. To some extend, this 
theory is less appropriate than three-body calculations to the problem 
of $\Delta B_{\Lambda\Lambda}$, since it replaces the basic two-body 
interactions by coupling to meson fields. However, our aim is to study 
the possibility to put the RMF on reasonable grounds, since a sensible 
estimate of $\Delta B_{\Lambda\Lambda}$ has implications on the 
calculations of multi-$\Lambda$ systems within this model.

The RMF approach 
has been  applied to hypernuclear systems containing
various amounts of hyperons (see for instance \cite{ruf,mz,schaf2}).
While the original $\sigma-\omega$ model well reproduces spectra of single 
$\Lambda$ hypernuclei (in particular, small spin-orbit splitting) 
its straightforward application to bound systems with two $\Lambda$
particles failed. It has been found
that the $\sigma-\omega$ model cannot provide a sufficiently attractive
$\Lambda\Lambda$ interaction and, consequently, the binding energy of
double-$\Lambda$ hypernuclei is substantially underestimated.
Considering the Lorentz-tensor $\Lambda$-$\omega$
coupling \cite{schaf1}, which allows for stronger couplings of $\Lambda$ to
mesons,
could lead to a stronger $\Lambda\Lambda$ interaction.
Detailed calculations revealed, however, that stronger $\Lambda$-meson
couplings do not necessarily result in a larger value of
$\Delta B_{\Lambda\Lambda}$\cite{jm1}.
Table~1 displays a typical example of RMF results for coupling constant ratios
$\alpha_{\omega}={{g_{\omega \Lambda}}\over{g_{\omega N}}}=1/3$ and 2/3,
compared to the empirical data.
 Quite recently, Schaffner et al\cite{schaf2} proposed to
strengthen the $\Lambda\Lambda$ binding by introducing an additional $YY$ 
interaction mediated
by two strange mesons (scalar $\sigma^*$ and vector $\Phi$) that couple 
exclusively
to hyperons. The coupling of hyperons to the $\Phi$ mesons was assumed to
satisfy the SU(6) relations,
whereas the coupling to $\sigma^*$ was fixed by
fitting to the estimated potential well depth for the $\Lambda$ hyperon in
a medium of other $\Lambda$ hyperons, $U_{\Lambda}^{(\Lambda)}\approx 20$~MeV.
The improved RMF model ($\sigma+\omega+\sigma^* + \Phi$) increases
the value $\Delta B_{\Lambda\Lambda}(^6_{\Lambda\Lambda}He)$ to about
3~MeV (from the original $<1$~MeV).

It is clear, however, that the discrepancy between the calculated and empirical
values
of $\Delta B_{\Lambda\Lambda}$ cannot be attributed entirely to the missing
meson exchanges between the two hyperons. To support this argument,  we
display in Table 2
\begin{equation}
\Delta S = S_{2n}-2S_{n}\;\;\; ,
\end{equation}
calculated for ordinary nuclei with two nucleons outside a spherical
core. Here $S_{n}$ ($S_{2n}$) is the separation
energy of the nucleon (2 nucleons). (Note that unlike the binding energies B
in case of hypernuclei, separation energies S are used in nuclear mass 
tables \cite{data}.) The RMF results are compared to
the experimental data \cite{data}. The model again  
fails to give the observed $\Delta S$.
In this case, possible additional mesons would have to be introduced 
from the beginning (fit to nuclear matter data) and thus are not 
expected to change the results.
Comparing the RMF and experimental values from Table 2 it is reasonable to 
expect a sizable contribution to $\Delta B_{\Lambda\Lambda}$ coming 
from other effects neglected in the mean field approximation \cite{ring}.
In the present work,
we aim to develop a simple model, which allows us to include the
two-$\Lambda$ correlation energy. This is a first step towards estimating
what fraction is left for the contribution
from $\sigma^*$ and $\Phi$ mesons. In this way, we are able to
extract information about their couplings to the $\Lambda$ hyperon.

It has to be stressed that the present work neglects $\Lambda$N 
correlations, which could lower the coupling constants and thus affect 
the $\Lambda\Lambda$ correlation energy. Similarly, the lack of spin 
dependence of the $\Lambda$-nucleus potential derived from the RMF may 
be of importance. We will address these questions in section III, where we
show on simple estimates that the present model yields a reasonable limit 
to the $\Lambda\Lambda$ correlation energy. 

The model used in our study is
described in the following section. In section 3, we present results of the
calculations. We show that the correlation energy between two $\Lambda$'s
is very sensitive to the RMF parametrization used.
Though for quark model inspired values of $\Lambda$ couplings
the correlation energy  contributes substantionally to
$\Delta B_{\Lambda\Lambda}$ it is not sufficient to account for the
empirical values. Indeed, extra meson exchanges proposed by Schaffner et al,
specific to the hyperon ($\Lambda$) sector are required.
Taking SU(6) value for the $\Phi \Lambda$ coupling constant $g_{\Phi\Lambda}$
we determine the $\sigma^*\Lambda$ coupling $g_{\sigma^*\Lambda}$ by
fitting the experimental data from observed 2$\Lambda$ hypernuclei.
Predictions are then made for $\Delta B_{\Lambda\Lambda}$ in heavier
hypernuclear systems. Conclusions are drawn in section 4.


\section{The Model}

A starting point to describe the relativistic system of
two interacting particles inside the nuclear medium would be
either the
covariant Bethe-Salpeter equation \cite{bs} or the manifestly covariant
formalism with constraints \cite{sad}.
However, due to the complexity of these approaches,
most of the works related to this subject are based on the
phenomenological equal-time two-body Dirac equation \cite{br,gf,scs,mp}.
Although not fully covariant, it has proved very useful in
understanding two-electron atoms and two-quark systems.
We will therefore adopt this approach here, as well.

In order to write down the Dirac equation for two $\Lambda$ particles in a
hypernucleus, we consider these two hyperons moving
in the scalar ($\sigma$) and vector ($\omega_0$) mean fields
brought about by the nucleons.
In addition, the hyperons interact with each other. Let's suppose that the
interaction is mediated by the exchange of the scalar
$(\sigma_\Lambda)$ and vector $(\omega_{0\Lambda})$ 
mesons (for simplicity, we neglect all the
other possible meson exchanges).
In accordance with the above assumptions the time-independent Dirac equation we
propose, neglecting retardation effects, has the following form:

\begin{eqnarray}\label{Ddl}
\big[-i\vec \alpha_1 . \vec \nabla_1 -i\vec \alpha_2 . \vec \nabla_2
+\beta_1(M_\Lambda +\Sigma(\vec r_1))+\beta_2(M_\Lambda +\Sigma(\vec r_2))&
\nonumber \\ [6mm]
+\beta_1\beta_2\Sigma_{S\Lambda}(\vec r_1,\vec r_2)
+\Sigma_{0\Lambda}(\vec r_1,\vec r_2)\big]\ \varphi_{\Lambda} (\vec r_1,\vec r_2)
&=2E_\Lambda \varphi_{\Lambda} (\vec r_1,\vec r_2),
\end{eqnarray}

\noindent
where
$\Sigma(\vec r_i)=\Sigma_S(\vec r_i)+\beta_i\Sigma_0(\vec r_i)$
represents the self-energy of a $\Lambda$ particle due to its interaction
with the nucleon fields.
The scalar ($\Sigma_S$) and time-like part of the vector interaction
($\Sigma_0$) are given by
$\Sigma_S(\vec r_i)=g_{\sigma \Lambda}\sigma (\vec r_i)$ and
$\Sigma_0(\vec r_i)=g_{\omega \Lambda}\omega_0 (\vec r_i)$, respectively.
$\Sigma_{S\Lambda}$ and $\Sigma_{0\Lambda}$ are the scalar and time-like vector
self-energies of the $\Lambda$ particles due to their mutual interaction.
In terms of $\sigma_\Lambda$ and $\omega_{0\Lambda}$ fields, they read:
$\Sigma_{S\Lambda}(\vec r_1,\vec r_2)=g_{\sigma \Lambda}\sigma_\Lambda(\vec r_1,\vec r_2)$,
$\Sigma_{0\Lambda}(\vec r_1,\vec r_2)=g_{\omega \Lambda}\omega_{0\Lambda}(\vec r_1,\vec r_2)$. Finaly,
$\varphi_{\Lambda} (\vec r_1,\vec r_2)$ is a 16-component spinor (labelled by two
indices), representing the two-$\Lambda$ state. $\alpha_i$ and $\beta_i$ ($i=$1,2)
are the Dirac matrices acting on the i-th spinor index.

\noindent
The fields $\sigma(\vec r)$ and $\omega_0(\vec r)$
fulfil the Klein-Gordon equations with the nuclear scalar and vector
densities as the source terms

\begin{equation}
 (\Delta -m_\sigma^2)\sigma=
g_{\sigma}\sum_{i=1,N}\bar \varphi_{N i}(\vec r) \varphi_{N i}(\vec r), \end{equation}

\begin{equation}
(\Delta -m_\omega^2)\omega_0
=-g_{\omega}\sum_{i=1,N}\bar \varphi_{N i}(\vec r)\gamma_0\varphi_{N i}(\vec r)\;\; . 
\end{equation}
Here, the spinors $\varphi_{N i}(\vec r)$ represent the one-nucleon states.

As stated above, each $\Lambda$ particle in addition moves in the
$\sigma_\Lambda$  and $\omega_{0\Lambda}$ fields whose source is the second hyperon.
In the approximation of heavy, static baryons the corresponding
Klein-Gordon equations for $\sigma_\Lambda$  and $\omega_{0\Lambda}$ acquire the form:

\begin{equation}
(\Delta -m_\sigma^2)\sigma_\Lambda=g_{\sigma \Lambda} \delta(\vec r_1-\vec r_2), \end{equation}

\begin{equation}
(\Delta -m_\omega^2)\omega_{0\Lambda}=
-g_{\omega \Lambda}\delta(\vec r_1-\vec r_2)
\;\; .\end{equation}

If the $\Lambda\Lambda$ interaction is neglected the Dirac equation (1)
reduces to two identical Dirac equations, each being equivalent to
the mean field approximation for a $\Lambda$ particle in a
hypernucleus.

To proceed further we express the two-$\Lambda$ spinor in terms of its 4 large
($\psi_\Lambda$), 8 medium ($\theta_\Lambda$, $\vartheta_\Lambda$) and 4 small ($\chi_\Lambda$)
components \cite{gf,scs}:

\begin{equation}
\varphi_\Lambda (\vec r_1,\vec r_2)={\psi_\Lambda \;\;\; \vartheta_\Lambda \choose
\theta_\Lambda \;\;\; \chi_\Lambda }
\end{equation}

If $\varphi_\Lambda$ from eq.(8) is brought into the Dirac equation (3)
and the components $\theta_\Lambda$, $\vartheta_\Lambda$, and $\chi_\Lambda$ eliminated,
the following equation for the components $\psi_\Lambda$ is obtained:

\begin{eqnarray}
\big[-\vec \sigma_1 . \vec \nabla_1{1 \over 2\bar M_{1\Lambda}}\vec \sigma_1 .\vec \nabla_1
      -\vec \sigma_2 . \vec \nabla_2{1 \over 2\bar M_{2\Lambda}}\vec \sigma_2 .\vec \nabla_2
+\Sigma_S(\vec r_1)+\Sigma_S(\vec r_2)&
\nonumber \\ [6mm]
+\Sigma_0(\vec r_1)+\Sigma_0(\vec r_2)
+\Sigma_{S\Lambda}(r)+\Sigma_{0\Lambda}(r)\big] \psi_\Lambda (\vec r_1,\vec r_2)
&=2\epsilon_\Lambda \psi_\Lambda (\vec r_1,\vec r_2),
\end{eqnarray}

\noindent
where $2 \bar M_{i\Lambda} =E^*_i+M^*_i-\Sigma_{0\Lambda}+\Sigma_{S\Lambda}$,
with $E^*_i=E_\Lambda-\Sigma_0(\vec r_i)$, $M^*_i=M_\Lambda+\Sigma_S(\vec
r_i)$, $\epsilon_\Lambda=E_\Lambda-M_\Lambda$ and $r$ is the relative distance between
two $\Lambda$ particles, $r=\vert \vec r_1-\vec r_2\vert$.

The equation (9) for $\psi_\Lambda (\vec r_1,\vec r_2)$ is still rather
complicated because $\bar M_{i\Lambda}$ depends on
$\vec r_1$ and $\vec r_2$. We simplify
the solution by neglecting the radial dependence of $\bar M_{i\Lambda}$ and
 replacing ${1 \over \bar M_{i\Lambda}}$ by the ground state expectation value
$<{1 \over \bar M_{i\Lambda}}>$.
This approximation leads to neglecting the spin-orbit interaction and terms
that renormalize somewhat the central potential.
However, since the spin-orbit interaction is very small for the
$\Lambda$ hyperon this does not represent a serious drawback.
The resulting (Schr\"{o}dinger type) equation for $\psi_\Lambda$ is then

\begin{eqnarray}
\big[-<{1 \over 2\bar M_{1\Lambda}}>{\vec \nabla_1}^2
     -<{1 \over 2\bar M_{2\Lambda}}>{\vec \nabla_2}^2
+\Sigma_S(\vec r_1)+\Sigma_S(\vec r_2) &
\nonumber \\ [6mm]
\label{Sdl}
+\Sigma_0(\vec r_1)+\Sigma_0(\vec r_2)
+\Sigma_{S\Lambda}(r)+\Sigma_{0\Lambda}(r)
\big]\psi_\Lambda (\vec r_1,\vec r_2)
&=\epsilon_\Lambda \psi_\Lambda (\vec r_1,\vec r_2)\;\;\; .
\end{eqnarray}

Although at this stage, equation (10) can be solved by expanding
$\psi_\Lambda (\vec r_1,\vec r_2)$ in a convenient basis, it is still tedious
enough that it is useful to look for further simplifications.

The $\Lambda$ particle in a hypernucleus spends most of its time in a
high density region, where the potential
$\Sigma_S(\vec r_i)+\Sigma_0(\vec r_i)$ can be approximated rather
accurately by a spherical harmonic oscillator
$W(\vec r_i)=-W_0+{1\over 2}M_\Lambda \omega^2r_i^2$ \cite{lm}. Consequently,
we shall use both that the RMF reproduces the hypernuclear spectra with
a great accuracy \cite{lmm} and that the potential seen by the $\Lambda$ is
very close to the hamonic oscillator to get a practical solution of (10).
Note that very similar approximations have been used in non-relativistic
calculations.

The $\sigma_{\Lambda}$ and $\omega_{\Lambda}$ exchanges
between $\Lambda$ hyperons (eqs.~6 and 7)
give rise to an effective $\Lambda$-$\Lambda$
potential $U(r)\equiv\Sigma_{S\Lambda}(r)+\Sigma_{0\Lambda}(r)$ ($r = |\vec r_1 -
\vec r_2|$) which reduces to a difference of two Yukawa forms:

\begin{equation}
U(r)\equiv\Sigma_{S\Lambda}(r)+\Sigma_{0\Lambda}(r)
=-{g_{\sigma\Lambda}^{2}\over {4\pi}}{e^{-m_\sigma r} \over r}+
{g_{\omega\Lambda}^{2}\over {4\pi}}{e^{-m_\omega r} \over r}\;\;\; .
\end{equation}

Above two approximations lead to  the
following replacement in the Schr\"{o}dinger-like equation (10):

\begin{equation}
\Sigma_S(\vec r_1)+\Sigma_S(\vec r_2)+\Sigma_0(\vec r_1)+\Sigma_0(\vec r_2)
+\Sigma_{S\Lambda}(r)+\Sigma_{0\Lambda}(r)
\to V(\vec r_1,\vec r_2)=W(r_1)+W(r_2)+U(r)\; .
\end{equation}

In fact, the oscillator depth $W_0$ and its frequency $\omega$ were not determined
by fitting $W(r_i)$ to $\Sigma_S(\vec r_i)+\Sigma_0(\vec r_i)$ 
but directly
fitted to the experimental energy spectrum of the particular $\Lambda$
hypernucleus. The two coupling constants $g_{\sigma\Lambda}$ and $g_{\omega\Lambda}$
from eq.(11) were chosen
to reproduce the spectroscopic data in the relativistic mean field
formalism \cite{lmm}, as accurately as possible, in the
whole ensemble of single-$\Lambda$ hypernuclei known. With the above determined
parameters, eq. (10) allows
us to estimate the correlation energy  of the two hyperons, which is neglected
in the mean field approximation.

Parameterizing the $\Lambda$ self-energies $\Sigma_S(\vec r_i)+
\Sigma_0(\vec r_i)$ in terms of HO potentials enables to express the
equation of motion (10) in Jacobi coordinates ($\vec R$, $\vec r$),
$$\vec r_1 = {1 \over \sqrt{2}}(\vec R + \vec r) \,\, ; \;\;\;
\vec r_2 = {1 \over \sqrt{2}}(\vec R - \vec r)\;\;\; , $$
and separate the centre of mass coordinates from the relative ones.
After straightforward manipulations the former equation (10) transforms
into the following two equations:

\begin{equation}
\left[ {\vec{P}^2 \over {2m_{\Lambda}}} + {1 \over 2} m_{\Lambda}\omega^2 
R^2 +2W_0 \right] \psi_R(R) = E_R \psi_R(R)\;\;\; ,
\end{equation}
\begin{equation}
\left[ {\vec{p}^2 \over {2m_{\Lambda}}} + {1 \over 2} m_{\Lambda}\omega^2 
r^2 +U(\sqrt{2}r) \right] \psi_r(r) = E_r \psi_r(r)\;\;\; ,
\end{equation}
where $\vec P$, $\vec p$ are the Jacobi impulse operators and 
$ m_{\Lambda}^{-1} = <{1 \over \bar M_{\Lambda}}> $.
  
If the harmonic oscillator parameters are fitted to  
eigenvalues of a single $\Lambda$ hypernucleus,
(taken either from RMF or from experiments) these two coupled 
equations yield a first approximation to the two $\Lambda$ 
binding energy  $B_{\Lambda \Lambda} \simeq - ( E_R + E_r)$. The correct 
value has to include at least two corrections :  a modification of the 
harmonic oscillator parameters due to the additional $\Lambda$ and the 
increase of the core energy, the so-called re-arrangement energy. 

The re-arrangement energy can be estimated in the RMF approximation as a
difference between the  $\Lambda$ eigenvalue and binding energy
\begin{equation} \Delta E_{core} = - E_{\Lambda} - B_{\Lambda} \ \ .
\end{equation}

The modification of the harmonic oscillator parameters due to the second
$\Lambda$ can be neglected as it is expected to be negligible in comparison with the above
$\Delta E_{core}$. This is because, whereas all the core 
particles contribute to $\Delta E_{core}$, only the added $\Lambda$ is 
affected by the change of the harmonic oscillator parameters.  

Furthermore, one can expect that the re-arrangement energy of 
the 2 $\Lambda$ hypernucleus is approximately twice the one of 
the single $\Lambda$ hypernucleus. 
Consequently, we obtain 

\begin{equation}
\Delta B_{\Lambda \Lambda} =  B_{\Lambda \Lambda} - 2 B_{\Lambda} 
\simeq -(E_R + E_r) + E_R + E_r(U=0) = E_r(U=0) - E_r  \ \ . \end{equation}

The alternative and equivalent way to determine $\Delta B_{\Lambda \Lambda}$
is to fit the harmonic oscillator parameters in such a way that the $\Lambda$
binding energy $B_{\Lambda} =-{1 \over 2} [E_{R} +E_{r}(U=0)]$ in  the
corresponding hypernucleus reproduces the empirical binding energies. Now
$B_{\Lambda}$  immediately incorporates $\Delta E_{core}$, and
similarly $B_{\Lambda\Lambda} = -(E_{R} +E_{r})$ includes the
re-arrangement of the core caused by the two $\Lambda$ particles 
$\simeq 2 \Delta E_{core}$. As a result, relation (16) is fulfilled again.

\section{The Results}

The model presented in the previous section was applied to calculation
of $\Delta B_{\Lambda\Lambda}$ for the following sample of double
$\Lambda$ hypernuclei: $^{6}_{\Lambda\Lambda}He$,
$^{10}_{\Lambda\Lambda}Be$,
$^{13}_{\Lambda\Lambda}B$, $^{18}_{\Lambda\Lambda}O$,
$^{42}_{\Lambda\Lambda}Ca$, $^{92}_{\Lambda\Lambda}Zr$ and
$^{210}_{\Lambda\Lambda}Pb$, which includes the measured cases.

Fitting
the HO parameters $\hbar\omega$~ and $W_0$ requires two experimental
values. Starting from $^{13}_{\Lambda}C$~ they are given by the 1s and
1p $\Lambda$ binding energies. For the three lightest elements, where the 1p
level is unbound, we extrapolated the $sp$ splitting from the C and O region.

We used three different RMF models, namely HS model
of Horowitz and Serot \cite{hs}, and models L1 and L3 of Lee et al. \cite{lee}
The masses and meson-nucleon coupling constants of $\sigma$ and $\omega$
mesons are presented in Table 3.
Different parametrizations allowed us to study the dependence of
$\Delta B_{\Lambda\Lambda}$ on the mass of the $\sigma$ meson $m_{\sigma}$. One would
expect that the smaller values of $m_{\sigma}$ (t.e., model L3) will give
larger correlation energy and consequently larger $\Delta B_{\Lambda\Lambda}$.

The couplings of the $\Lambda$ hyperon to the meson fields are often 
defined via coupling constant ratios $\alpha_{i}={g_{i\Lambda}\over g_{iN}}$,
$i=\sigma, \omega$. For each of the above RMF parametrizations we used two
coupling ratios $\alpha_{\omega}= 1/3$ (a) and 2/3 (b). 
Whereas the value of 2/3 is predicted by the constituent quark model, 1/3
ratio was widely used in the pioneering RMF hypernuclear calculations. 
The corresponding $\alpha_{\sigma}$ was then chosen to fit the hypernuclear 
spectra \cite{mjc}.
The ratios $\alpha_{i}$ are included in the list of parameters in Table 3,
as well.

The $\Delta B_{\Lambda\Lambda}$ corresponding to the different parametrizations
of Table~3 are displayed in Table~4.
The results indicate that $\Delta B_{\Lambda\Lambda}$
depends on the model used. The values of  $\Delta B_{\Lambda\Lambda}$ are 
larger for lower values of $m_{\sigma}$ as predicted. In addition,
 $\Delta B_{\Lambda\Lambda}$ is quite sensitive to the coupling
ratios $\alpha_{\omega}$. Whereas for $\alpha_{\omega} = 1/3$ there is
hardly any improvement over the RMF values, 0.5 - 1.0 MeV is gained
with $\alpha_{\omega}=2/3$.

The results of Table~4 indicate also that including
the correlation energy from the
$\sigma$ and $\omega$ exchange, though sizable in the case of
$\alpha_{\omega} = 2/3$, cannot by itself
account for empirical $~4.5$~MeV of the $\Delta B_{\Lambda\Lambda}$ in
light hypernuclei.
Note that the results of Table~4, for light nuclei, cannot be compared
directly to those of Table~1 on a quantitative level, because of small
differences used in each calculation. Qualitatively, however, the strong
dependence of the correlation energy on the strength of the  
$\omega$-coupling and its large incidence on  $\Delta B_{\Lambda\Lambda}$ for 
2/3 can be taken for granted.
Therefore, according to the chosen parametrization,
at least half of the empirical $\Delta B_{\Lambda\Lambda}$ has to
come from "new" meson exchanges that are not included in the original 
versions of RMF models.

In order to investigate the range of coupling constants needed to bring
the calculated $\Delta B_{\Lambda \Lambda}$ into agreement with experiments,
we followed the work of Schaffner
et al \cite{schaf2} and assumed scalar  $\sigma^*$
and vector $\Phi$ meson fields.
We addopted their meson masses, namely $m_{\sigma^*} = 975.0$ MeV and
$m_{\Phi} = 1020.0$ MeV, respectively. Similarly the $\Phi$ coupling is taken
from the SU(6) relations, $\alpha_{\Phi}={g_{\Phi\Lambda}\over g_{\omega N}}
=-{\sqrt{2}\over 3}$. Contrary to their work, the $\sigma^*$ coupling
is considered as a free parameter to be fitted to the empirical
values of $\Delta B_{\Lambda \Lambda}$. Since $\sigma^*$ and $\Phi$ act only 
between two $\Lambda$'s, they simply modify the potential $U(r)$ to be used in 
(14). 

The calculated $\Delta B_{\Lambda\Lambda}$ as a function of $\alpha_{\sigma^*}$
are presented in Fig.~1 for $^6_{\Lambda \Lambda}He$,
$^{10}_{\Lambda \Lambda}Be$ and $^{13}_{\Lambda \Lambda}B$. Use is made
of the HS parametrization of the RMF, which stands roughly in between 
the two other cases $L_1$ and $L_3$ as for the magnitude of the predicted 
 $\Delta B_{\Lambda\Lambda}$. The two sets of curves corresponding to
$\alpha_{\omega} =$ 1/3 and 2/3 intercept the domain defined by the
experimental values at $\alpha_{\sigma^*}$ around 0.71 and 0.79,
respectively. 
In view of the large experimental errors (Table~1) these values are only
approximative.  Nevertheless, it means that inspite of the correlation 
effects taken into account, which reduce the short range repulsion effect, 
the largest repulsive $\alpha_{\omega}$ implies the largest 
attractive $\alpha_{\sigma^*}$ coupling.

Having fixed $\alpha_{\sigma^*}$ we performed the calculations for
the set of double $\Lambda$ hypernuclei mentioned above. It provides us
with the prediction of the A-dependence of $\Delta B_{\Lambda \Lambda}$.
The results are displayed in fig.~2.
We observed a decrease of the bond energy with A which is roughly the one
predicted by a crude perturbative estimate of the $\Lambda \Lambda$
interaction. This result confirms recent calculations by Lanskoy {\it 
et al} [11] based on a Skyrme-Hartree-Fock approach, which show a 
comparable decrease of $\Delta B_{\Lambda\Lambda}$ with A.

We shall end up this section by discussing two effects which could
qualitatively affect the present conclusions. The first one concerns the
spin dependence of the $\Lambda \Lambda $ potential. Whereas the two
$\Lambda  $ are in a singlet state, the actual determination of the
$\Lambda$ coupling constants relies on the $\Lambda N $ spin average. In other
words, the $\Lambda \Lambda$ interaction is somewhat underestimated, the
singlet potential being known to be more attractive than the triplet one.

In order to get an idea by how much this effect influences the value of the
coupling constants, we compared the singlet and triplet effective YNG
interactions of Yamamoto and Bando\cite{yb}. The ratio of their strengths was
then used to determine $V^{\Lambda N}_{singlet}$ from the spin average
$V^{\Lambda N}$ RMF interaction. We left the vector coupling unchanged and
modified the scalar coupling constant. The resulting $\alpha_{\sigma}$ relevant for
the singlet state increased by 2.8~\% and 4.3~\% for $\alpha_{\omega}$ = 2/3
and 1/3, respectively.

The second effect is acting in the opposite direction. Namely, the
$\Lambda N$ interaction determined from hypernuclei contains implicit
correlations, whereas the estimate of the $\Lambda \Lambda$ correlation
energy should rely on the bare potential. This last should be determined
from the RMF interaction by unfolding with an appropriate $\Lambda N$
correlation function.

To obtain at least a rough estimate of the effect we used the correlation
function of Pare\~{n}o et  al \cite{prb} and folded the U(r) 
interaction entering eq.~(14). The $\alpha_{\sigma}$ coupling constant 
ratio appearing in U(r) is then
decreased in order to get exactly the same eigenvalue of eq.~(14) as before. We
determine in this way the bare scalar coupling constant, $g_{\sigma
\Lambda}$ while $\alpha_{\omega}$ is kept unchanged. This procedure ends in
a decrease of $\alpha_{\sigma}$ by 6.3~\% and 1.7~\% for $\alpha_{\omega}$ = 2/3
and 1/3, respectively. These results have been confirmed by a second
estimate based on a correlation function constructed from the approach
described in \cite{sm}, 
which leads to even slightly lower values. 

Adding the two effects we conclude that they tend to cancel each other to a
large extent. Consequently the RMF coupling constants might change up to 
4~\%. The uncertainty in the results of table 4 due to the neglecting of these
two effects are well within the approximations used in the present model.

\section{Conclusions.}

This paper is devoted to the binding energy of double $\Lambda$
hypernuclei, more precisely to the bond energy $\Delta
B_{\Lambda \Lambda}$ as defined
by (1). We show that within the relativistic mean field approach, part of this
energy is provided by the short range correlation, the remaining
being due to the exchange of $\sigma*$ and $\Phi$ mesons between the
two $\Lambda$s. The balance between the two effects depends sensitively
on the coupling of the $\Lambda$ to the $\omega$ field. Whereas for a
coupling constant $\alpha_{\omega}$ = 1/3  the correlation effects are
not very efficient, they become sizable at higher values, doubling the
RMF results for $\alpha_{\omega}$ = 2/3.

The present results have been obtained by reducing a relativistic two-body
equation of the Breit type to a Schr\"odinger equation. Furthermore,
advantage has been taken of the fact that the average potential
experienced by the $\Lambda$ in a nucleus is very close to an harmonic
oscillator potential. In this way the calculations are considerably
simplified.

Although more sophisticated calculations are desirable, they are not
expected to change the present results, at least at a semi-quatitative
level. We recall that for reasons stated in the introduction our 
estimate is an upper limit to the $\Lambda-\Lambda$ correlation energy.

We found that, according to the $\omega-\Lambda$ coupling, at least half of
$\Delta B_{\Lambda \Lambda}$ arises from the meson exchanges specific to the
$\Lambda\Lambda$ interaction. In such a case one may suspect the argument
advocated in the introduction, stating that for ordinary nuclei $\Delta S$
is essentially due to correlation effects. Actually, it is very easy to get
convinced from toy models  that the gain in binding energy coming from the
short-range two body correlation is dominated by the repulsive $\omega$
field. Indeed, assuming $\alpha_{\omega}=1.$ $\Delta B_{\Lambda \Lambda}$
gets close to 3.5~MeV. Thus, the difference between the $\Lambda$ and the
nucleon case reflects the strength of their coupling to the $\omega$ field. 

We remind the reader that the RMF theory cannot compete with more 
elaborated three-body (cluster) calculations of  $\Delta B_{\Lambda 
\Lambda}$. In particular for such a light system as $^{\; 6}_{\Lambda 
\Lambda} He$, its application is questionable. In view of extensions to 
multi-$\Lambda$ systems, however, it is important to check the 
constraints it brings on the coupling of the $\Lambda$ to the various 
meson fields. In this respect, it would be very desirable to obtain 
experimental data for heavier nuclear cores than those actually 
available. 

\vspace*{10mm}
{\it Acknowledgment:}
J.M. would like to thank for the hospitality of the Division of 
Theoretical Physics, IPN Orsay.  
J.M. would also like to acknowledge support from GACR grant number
202 0442.

\begin{table}
{\bf Table 1 --}
$B_{\Lambda\Lambda}$ and $\Delta B_{\Lambda\Lambda}$ (in MeV)
(see text for definition) of the double $\Lambda$-hypernuclei
$ ^{\; 6}_{\Lambda\Lambda}He $,
$^{10}_{\Lambda \Lambda} Be$ and $^{13}_{\Lambda \Lambda} B$.
 The RMF predictions for the HS parametrization $\cite{hs}$ with the 
 coupling constant ratios $\alpha_{\omega}={{g_{\omega\Lambda}}\over{g_{\omega N}}}=1/3$ (a)
and $\alpha_{\omega}=2/3$ (b) are compared with the experimental
values \cite{fran}.
\vspace{2mm}
\begin{center}
\begin{tabular} {|c|ccc|ccc|}
& \multicolumn{3}{c|}{$B_{\Lambda\Lambda}$} &
\multicolumn{3}{c|}{${\Delta}B_{\Lambda\Lambda}$}\\
\hline
 & a & b & EXP &  a & b & EXP\\
\hline
\hline
$^{\; 6}_{\Lambda \Lambda} He$ & 4.8 & 5.3
& $10.9 \pm 0.6$ &
 0.8 & 0.5 & $4.7 \pm 0.6$ \\ 
$^{10}_{\Lambda \Lambda} Be$
& 14.8 & 15.1 & $17.7 \pm 0.4$ &
 1.1 & 0.6 & $4.3\pm 0.4$ \\ 
$^{13}_{\Lambda \Lambda} B$
& 23.5 & 23.1 & $27.5 \pm 0.7$ &
 1.0 & 0.4 & $4.8 \pm 0.7$ \\
\end{tabular}
\end{center}
\end{table}

\begin{table}
{\bf Table 2 --}
Comparison of $\Delta S$ (in MeV) (see eq.(2) for definition)
calculated within the RMF model for the HS parametrization \cite{hs} 
with the experimental values \cite{data} for selected nuclei.
\vspace{2mm}
\begin{center}
\begin{tabular} {|c|cccccc|}
 & $^{18}O$ & $^{30}Si$ & $^{38}Ar$ & $^{42}Ca$ & $^{92}Zr$ & $^{210}Pb$ \\
\hline 
RMF & 0.55 & 0.21 & 0.29 & 0.27 & 0.13 & 0.11 \\
EXP & 3.90 & 2.14 & 3.05 & 3.12 & 1.44 & 1.25 \\
\end{tabular}
\end{center}
\end{table}

\begin{table}
{\bf Table 3 --}
The parametrizations used in this work. Meson masses (in MeV)
and meson-nucleon coupling constants were adopted from
refs.\cite{hs} (HS) and \cite{lee} (L1 and L3). Two sets (denoted by a) and b)) of
the coupling ratios $\alpha_i = {{g_{i\Lambda}}\over {g_{i N}}}$ 
($i=\sigma,\;\omega$) are presented, as
well.
\vspace{2mm}
\begin{center}
\begin{tabular} {|c|cccc|cc|}
 & $m_{\sigma}$ & $g_{\sigma N}$ & $m_{\omega}$ & $g_{\omega N}$
&\multicolumn{2}{c|}{ $\alpha_{\sigma}$ }   \\ \hline
& & & & & {\bf a} ($\alpha_{\omega}=1/3$) & {\bf b} ($\alpha_{\omega}=2/3$)\\
\hline \hline
HS & 520.0 & 10.481 & 783.0 & 13.814 
&  0.342 & 0.623 \\
L1 & 550.0 & 10.30  & 783.0 & 12.60  
& 0.334   & 0.607 \\
L3 & 492.26 & 10.692 & 780.0 & 14.8705 
& 0.341 & 0.624 \\
\end{tabular}
\end{center}
\end{table}

\begin{table}
{\bf Table 4 --}
$\Delta B_{\Lambda\Lambda}$ (in MeV) for the
$\sigma-\omega$ model with the parametrizations
of Table 3 as a function of the mass number.
\vspace{2mm}
\begin{center}
\begin{tabular}{|cccccccc|}
  &  He & Be & B & O & Ca & Zr & Pb \\
\hline 
HS  & & & & & & & \\
a & 0.96 & 0.93 & 0.91 & 0.78 & 0.55 & 0.43 & 0.20 \\
b & 1.82 & 1.69 & 1.60 & 1.41 & 1.01 & 0.79 & 0.38 \\
\hline
L1 & & & & & & &  \\
a & 0.78 & 0.76 & 0.74 & 0.63 & 0.44 & 0.34 & 0.16\\
b & 1.22 & 1.13 & 1.07 & 0.94 & 0.67 & 0.53 & 0.25 \\
\hline
L3 & & & & & & & \\
a & 1.05 & 1.02 & 0.99 & 0.86 & 0.61 & 0.48 & 0.23 \\
b & 2.33 & 2.17 & 2.05 & 1.82 & 1.31 & 1.04 & 0.51 \\
\end{tabular}
\end{center}
\end{table}

\begin{figure}

\caption{$\Delta B_{\Lambda\Lambda}$ for $^6_{\Lambda\Lambda}He$ (dotted line),
$^{10}_{\Lambda\Lambda}Be$ (dashed line), and $^{13}_{\Lambda\Lambda}B$
(solid line) as a function of
the coupling ratio $\alpha_{\sigma^*}$ calculated for two different
$\Lambda-\omega$ coupling ratios ($\alpha_{\omega}$=1/3 and 2/3) 
using HS parametrization. The horizontal dotted lines indicate the spreading
of experimental values without errors (see Table~1). 
}
\end{figure}

\begin{figure}
\caption{$\Delta B_{\Lambda\Lambda}$ as a function of A calculated 
within HS parametrization for two  different
$\Lambda-\omega$ coupling ratios: $\alpha_{\omega}$=1/3 (dashed line) and
$\alpha_{\omega}$=2/3 (solid line). Experimental values are also displayed. 
} 
\end{figure}
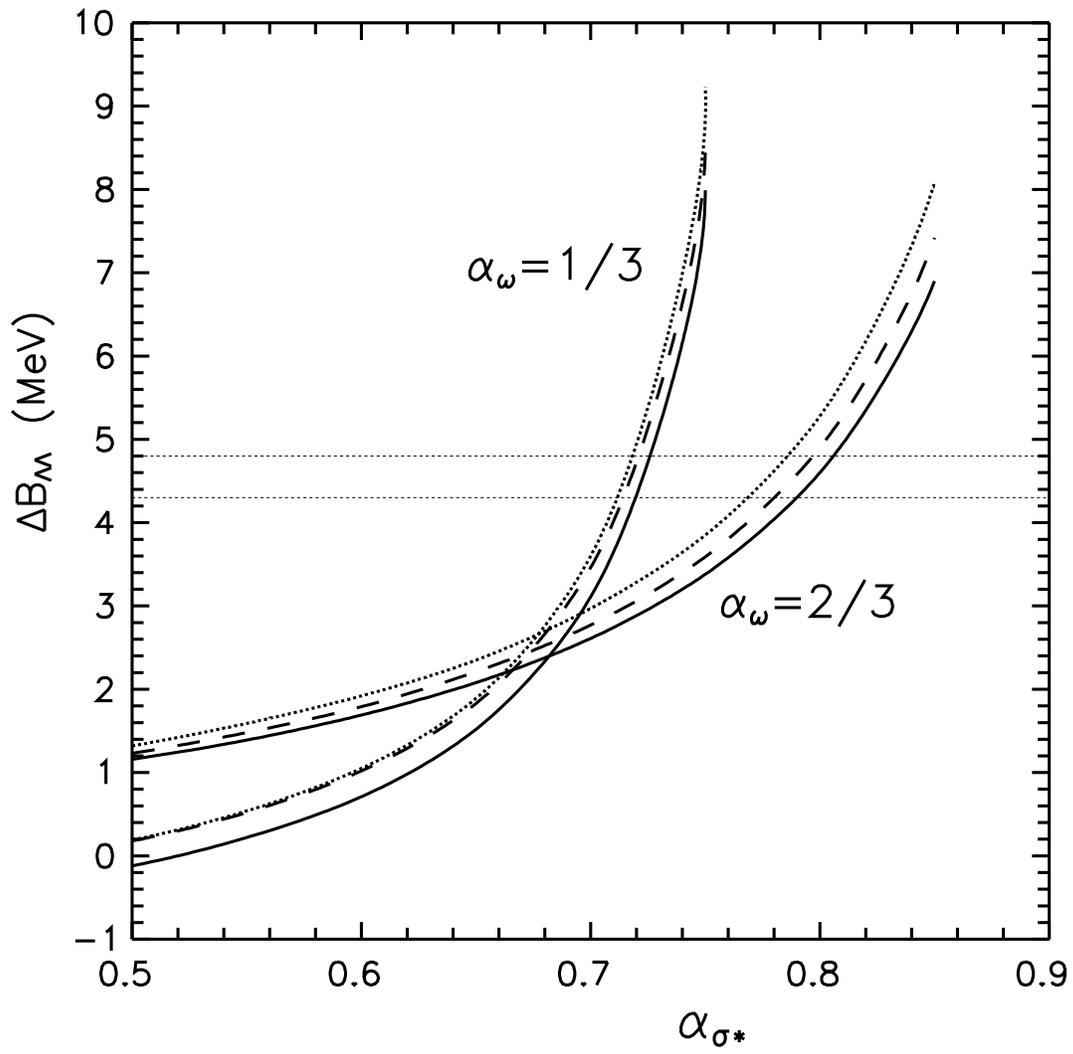
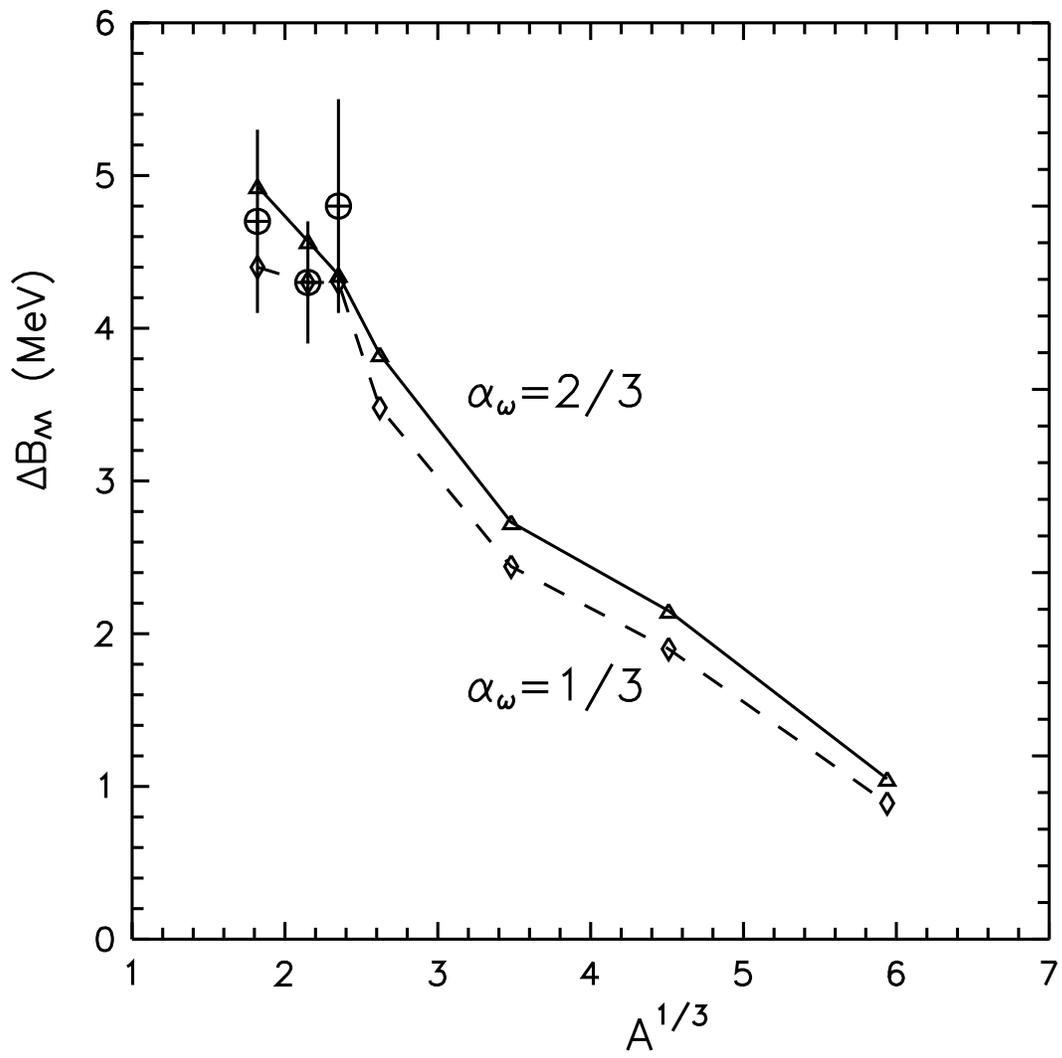
\end{document}